\documentclass[runningheads,a4paper]{llncs}

\usepackage{amsmath,amssymb,amscd}
\usepackage{graphicx}
\usepackage{tikz}

\usepackage[latin1]{inputenc}

\newcommand{\Go}[1]{{G\"{o}del} }

\usepackage{color}

\newenvironment{proof2}{\trivlist\item[\hskip
       \labelsep{\it Proof:\/}]\ignorespaces }{\hfill$\blacksquare $\endtrivlist}

\begin{document}

\date{}
\title{Possibilistic semantics for a modal $KD45$ extension of  G\"odel fuzzy logic}
\author{F\'elix Bou\inst{1} \and Francesc Esteva\inst{1} \and Llu\'{\i}s Godo\inst{1} \and  Ricardo \'Oscar Rodr\'iguez\inst{2}}

\institute{Artificial Intelligence Research Institute, IIIA - CSIC, Bellaterra, Spain \ \email{\{fbou,esteva,godo\}@iiia.csic.es}
\and
Departamento de Computaci\'on, FCEyN - UBA, Argentina \\
\email{ricardo@dc.uba.ar}}
\maketitle

\begin{abstract}
In this paper we  provide a simplified semantics for the logic $KD45(\mathbf{G})$, i.e. the many-valued G\"odel counterpart  of the classical modal logic $KD45$. More precisely, we characterize $KD45(\mathbf{G})$ as the set of valid formulae of the class of possibilistic G\"odel Kripke Frames $\langle W, \pi \rangle$,
where $W$ is a non-empty set of worlds and $\pi: W \to [0, 1]$ is a normalized possibility distribution on $W$.
\end{abstract}

\section{Introduction}

 {\em Possibilistic logic} \cite{DuLaPr,DuPr2} is a well-known uncertainty logic  to reasoning with graded beliefs on classical propositions by means of necessity and possiblity measures.
 These measures are defined in terms of possibility distributions. A (normalized) possibility distribution is a mapping  $\pi: \Omega \to [0, 1]$, with $\sup_{w \in \Omega} \pi(w) = 1$, on the set $\Omega$ of classical interpretations of a given propositional language that ranks interpretations according to its plausibility level: $\pi(w) = 0$ means that $w$ is rejected, $\pi(w) = 1$ means that $w$ is fully plausible, while $\pi(w) < \pi(w')$ means that $w'$ is more plausible than $w$.
 A possibility distribution $\pi$ induces a pair of dual possibility and necessity measures on propositions, defined respectively as:
\begin{center}
$\Pi(\varphi) = \sup\{ \pi(w) \mid w \in \Omega, w(\varphi) = 1 \}$ \\
$N(\varphi) = \inf\{ 1- \pi(w) \mid w \in \Omega, w(\varphi) = 0\}$ .
\end{center}
From a logical point of view, possibilistic logic can be seen as a sort of graded extension of the non-nested fragment of the well-known modal logic of belief $KD45$ \cite{DuPrSc}, in fact, $\{0, 1\}$-valued possibility and necessity measures over classical propositions can be taken as equivalent semantics for the modal operators of the logic $KD45$ \cite{BD13}.

When trying to extend the possibilistic belief model  beyond the classical framework of Boolean propositions to many-valued propositions,  one has to come up with appropriate extensions of the notion of necessity and possibility measures for them (see e.g.  \cite{DeGoMa}). In the particular context of G\"odel fuzzy logic \cite{Hajek98}, natural generalizations that we will consider in this paper are the following. 
A possibility distribution $\pi: \Omega \to [0, 1]$ on the set $\Omega$ of G\"odel propositional interpretations induces the following generalized possibility and necessity measures over G\"odel logic propositions:
\begin{itemize}

\item[] $\Pi(\varphi) = \sup_{w \in \Omega} \{ \min(\pi(w),  w(\varphi)) \}$

   $N( \varphi) = \inf_{w \in \Omega}  \{\pi(w) \Rightarrow w(\varphi)  \}$
\end{itemize}
where $\Rightarrow$ is G\"odel implication, that is, for each $x, y \in [0, 1]$, $x \Rightarrow y = 1$ if $x \leq y$,  $x \Rightarrow y = y$, otherwise.\footnote{Strictly speaking, the possibility measure is indeed a generalization of the classical one, but the necessity measure is not, since $x \Rightarrow 0 \neq 1-x$.}
These expressions agree with the ones commonly used in many-valued modal Kripke frames $(W, R)$ to respectively evaluate modal formulas $\Diamond \varphi$ and $\Box \varphi$ (see for example \cite{BouJLC} and references therein) when the $[0, 1]$-valued accessibility relation $R: W \times W \to [0, 1]$
is  defined by a possibility distribution $\pi: W \to [0, 1]$ as $R(w,w') = \pi(w')$, for any $w, w' \in W$.

Actually, modal extensions of G\"odel fuzzy logic have been studied by Caicedo and Rodr\'iguez \cite{CaiRod2015}, providing sound and complete axiomatizations for different classes of
$[0, 1]$-valued Kripke models. These structures are triples $M = (W, R, e)$, where $W$ is a set of worlds, $R = W \times W \to [0, 1]$ is a many-valued accessibility relation and $e: W \times Var \to [0, 1]$ is such that, for every $w \in W$, $e(w, \cdot)$  is a G\"odel $[0,1]$-valued evaluation of propositional variables (more details in next section) that extends to modal formulas as follows:

\begin{itemize}

\item[]  $e(w,\Diamond \varphi )=\sup_{w'\in W}\{\min(R(w,w'), e(w',\varphi ))\}$.

$e(w,\Box \varphi )=\inf_{w'\in W}\{R(w,w')\Rightarrow e(w',\varphi )\}$

\end{itemize}
We will denote by 
${\cal KD}45(G)$ the class of $[0, 1]$-models $M = (W, R, e)$ where $R$ satisfies the following many-valued counterpart of the classical properties:

\begin{itemize}
\item{\em Seriality}: $\forall w \in W$, $\sup_{w' \in W} R(w, w') = 1$.
\item {\em Transitivity}: $\forall w, w', w'' \in W$,  $\min(R(w, w'), R(w', w'')) \leq R(w, w'')$
\item {\em Euclidean}: $\forall w, w', w'' \in W$,  $\min(R(w, w'), R(w, w'')) \leq R(w', w'')$
\end{itemize}

In this setting,  the class $\Pi\mathcal{G}$ of {\em possibilistic Kripke models} $(W, \pi, e)$, where $\pi: W \to [0, 1]$ is a normalized possibility distribution on the set of worlds $W$, can be considered as the subclass of ${\cal KD}45(G)$ models $(W, R, e)$ where $R$ is such that $R(w, w') = \pi(w')$.  Since  $\Pi\mathcal{G} \subsetneq {\cal KD}45(G)$, it follows that the set $Val({\cal KD}45(G))$ of valid formulas in the class of ${\cal KD}45(G)$ is a subset of the set $Val(\Pi\mathcal{G} )$ of valid formulas in the class $\Pi\mathcal{G}$, i.e.
$Val({\cal KD}45(G)) \subseteq Val(\Pi\mathcal{G} )$.

In the classical case (where  truth-evaluations, accessibility relations
and possibility distributions are $\{0, 1\}$-valued) it is well known that (see e.g. \cite{Pietrus}) 
 that the semantics provided by the
class of Kripke frames with serial, transitive and euclidean
accessibility relations is  equivalent to the class of Kripke frames
with semi-universal accessibility relations (that is, relations of the
form $R = W \times E$, where $\emptyset \neq E \subseteq W$). But the
latter models are nothing else than $\{0, 1\}$-valued possibilistic models, given by the characteristic functions of the $E$'s.

However, over G\"odel logic, the question of whether the semantics provided by the
class of $[0, 1]$-valued serial, transitive and euclidean Kripke frames  is equivalent to the possibilistic semantics, that is, whether $Val(\Pi\mathcal{G}) = Val( {\cal KD}45(G))$ also holds, is not known. In this paper we positively solve this  problem. Indeed  we show that Caicedo-Rodriguez's  G\"odel modal logic $KD45(\mathbf{G})$ \cite{CaiRod2015} properly captures the above  possibilistic semantics. In this way, we  extend  the results obtained in \cite{DeGoMa} for the non-nested fragment of the modal language. We also note that this problem has already been solved for logics over finite and linearly ordered residuated lattices (MTL chains), thus in particular for finite-valued G\"odel logics, but with a language expanded with truth-constants and with Baaz-Monteiro operator $\Delta$, see \cite{HHEGG94,BouEstGo}. 

After this  introduction,  in the next section we first summarize the main results by Caicedo-Rodriguez on G\"odel modal logic $KD45(\mathbf{G})$ and its semantics given by $[0, 1]$-valued serial, transitive and euclidean Kripke models. Then we consider our many-valued possibilistic Kripke semantics, and prove in the last section that it is equivalent to the relational one. We conclude with some open questions that we leave as future research. We also include an appendix with several technical proofs.

\section{G\"odel Kripke Frames}
In their paper \cite{CaiRod2015} Caicedo and Rodr\'iguez consider a modal logic over G\"odel logic.  The language $\mathcal{L}_{\square \Diamond }(Var)$ of propositional \emph{bi-modal logic} is built from a countable set $Var$ of
propositional variables, connectives symbols $\vee ,\wedge ,\rightarrow
,\bot ,$ and the modal operator symbols $\square $ and $\Diamond $. We
will simply write $\mathcal{L}_{\square \Diamond }$ assuming $Var$ is known and fixed..
Then, the modal semantics is defined as follows. 

\begin{definition}
\label{godelkripkeframe} \emph{A} G\"odel-Kripke frame ($GK$-frame)\ \emph{will be a
structure }$\mathcal{F}=\langle W,R\rangle $\emph{\ where }$W$\emph{\ is a non-empty
set of objects that we call }worlds of\emph{\ }$\mathcal{F},$\emph{\ and }$R:W\times
W \to \lbrack 0,1]$.
A $\mathcal{F}$-Kripke G\"odel model is a pair $M=\langle \mathcal{F},e\rangle$ where $\mathcal{F}$ is a GK-frame and $e:W\times Var\rightarrow \lbrack 0,1]$ provides in each world an evaluation of variables. $e$ is inductively extended to arbitrary formulas in the following way:  \vspace{0.1cm}

\begin{tabular}{ ll}
$e(w,\varphi \wedge \psi ) = \min(e(w,\varphi ), e(w,\psi ))$ \mbox{ }
&
$e(w,\varphi \vee \psi ) = \max(e(w,\varphi ),  e(w,\psi ))$ \\

$e(w,\varphi \rightarrow \psi ) =e(w,\varphi )\Rightarrow e(w,\psi )$
& 
$e(w,\bot ) =0$ 
\end{tabular}

\begin{tabular}{l}
 $e(w,\Box \varphi ) =\inf_{w' \in W}\{R(w,w')\Rightarrow e(w',\varphi )\}$ 
\\
$e(w,\Diamond \varphi )=\sup_{w'\in W}\{\min(R(w,w'), e(w',\varphi ))\}$.
\end{tabular}
\end{definition}

\noindent Truth, validity and entailment are defined as usual: given a GK-model $M = (W, R, e)$, we write $(M, w) \models \varphi$ when $e(w, \varphi) = 1$, and $M \models \varphi$ if $(M, w) \models \varphi$ for every $w \in W$; given a class of GK-models ${\cal N}$, and a set of formulas $T$, we write $T \models_{\cal N} \varphi$ if, for every model $M =(W, R, e)$ and $w \in W$, $(M, w) \models \varphi$ whenever $(M, w) \models \psi$ for every $\psi \in T$.

In \cite{CaiRod2015} it is shown that the set $Val({\cal K}) = \{ \varphi \mid \ \models_{\mathcal K} \varphi \}$ of valid formulas in $\cal K$,  the class of all GK-frames,  is axiomatized by adding the following additional axioms and rule to those of G\"odel fuzzy logic G (see e.g. \cite{Hajek98}):  \\

\begin{tabular}{rlrl}
$(K_\Box)$ & $\Box(\varphi \to \psi) \to (\Box \varphi \to \Box \psi)$ &
$(K_\Diamond)$ & $\Diamond(\varphi \lor \psi) \to (\Diamond\varphi  \lor \Diamond \psi)$\\
$(F_\Box)$ & $\Box \top$ &
$(P)$ & $\Box(\varphi \to \psi) \to (\Diamond \varphi \to \Diamond \psi)$\\
$(FS2)$ & $(\Diamond\varphi  \to \Box \psi) \to \Box(\varphi \to \psi)$ \mbox{} \mbox{} \quad & 
$(Nec)$ & from $\varphi$ infer $\Box \varphi$\\
\end{tabular}
\vspace{0.2cm}

\noindent The resulting logic will be denoted $K(G)$. Moreover, in  \cite{CaiRod2015} it is also shown that the set $Val( {\cal KD}45(G))$ of valid formulas in the subclass  of GK-models ${\cal KD}45(G)$ is axiomatized by adding the following additional axioms: \vspace{0.2cm}

\begin{tabular}{ll@{\qquad}ll}
$(D)$ & $\Diamond \top$ & &  \\
$(4_\Box)$& $\Box \varphi \to \Box \Box \varphi$ &$(4_\Diamond)$& $\Diamond\Diamond \varphi \to  \Diamond \varphi$ \\
$(5_\Box)$&  $\Diamond\Box \varphi \to  \Box \varphi$ &$(5_\Diamond)$& $\Diamond \varphi \to \Box \Diamond \varphi$ \\
\end{tabular}
\vspace{0.2cm}

\noindent The  logic obtained by adding these axioms to $K(G)$ will be denoted $KD45(\mathbf{G})$.

\section{More about $KD45(\mathbf{G})$}

Let  $\vdash _{G}$
denote deduction in G\"odel fuzzy logic G. Let $\mathcal{L}(X)$ denote the set of
formulas built by means of the connectives $\wedge ,\rightarrow ,$ and $\bot
,$ from a given subset of variables $X \subseteq Var$. For simplicity, the extension of a valuation $%
v:X\rightarrow \lbrack 0,1]$ to $\mathcal{L}(X)$ according to G\"odel logic
interpretation of the connectives will be denoted $v$ as well. It is well
known that G is complete for validity with respect to these
valuations. We will need the fact that it is
actually sound and complete in the following stronger sense, see \cite{CaiRod2010}.

\begin{proposition} \label{ordersoundness}
\begin{itemize}
\item[i)] If $T\cup \{\varphi \}\subseteq \mathcal{L}(X)$, then
 $T\vdash _G \varphi $ iff $\inf v(T)\leq v(\varphi )$
for any valuation $v:X\rightarrow \lbrack 0,1]$.

\item[ii)]If $T$ is countable,
and $T\nvdash _ G \varphi _{i_{1}}\vee ..\vee \varphi _{i_{n}}$
for each finite subset of a countable family $\{\varphi _{i}\}_{i\in I}$ there is
an evaluation $v:\mathcal{L}(X) \rightarrow \lbrack 0,1]$\ such that $v(\theta )=1$\ for all
$\theta \in T$\ and $v(\varphi _{i})<1$ for all $i \in I$.

\end{itemize}
\end{proposition}

\noindent The following are some theorems of $K(G)$,  see \cite{CaiRod2015}: \vspace{0.2cm}

$%
\begin{array}{ll}
T1. & \lnot \Diamond \theta \leftrightarrow \square \lnot \theta
\\
T2. & \lnot \lnot \square \theta \rightarrow \square \lnot \lnot
\theta \\
T3. & \Diamond \lnot \lnot \varphi \rightarrow \lnot \lnot \Diamond
\varphi \\
T4. & (\square \varphi \rightarrow \Diamond \psi )\vee \square
((\varphi \rightarrow \psi )\rightarrow \psi )\\
T5. & \Diamond (\varphi \to \psi) \to (\square \varphi \to \Diamond \psi)
\end{array}%
$ \vspace{0.2cm}

\noindent The first
one is an axiom in Fitting's systems in \cite{Fitting91}, the next two were
introduced in \cite{CaiRod2015}, the fourth one will be useful in our completeness
proof and is the only one depending on prelinearity. The last is known as the first
connecting axiom given by Fischer Servi.

Next we show that in $KD45(\mathbf{G})$  iterated modalities can be simplified. This is in accordance with our intended possibilistic semantics for $KD45(\mathbf{G})$ that will be formally introduced in next section.

\begin{proposition} \label{simplif}
The logic $KD45(\mathbf{G})$ proves the following schemes:
\vspace{0.2cm}

$
\begin{array}{lll}
({F}_{\Box\Diamond}) && \Diamond\Box\top \leftrightarrow \Box \Diamond \top \leftrightarrow  \neg \bot \\
({U}_\Diamond)  & & \Diamond\Diamond  \varphi \leftrightarrow  \Diamond  \varphi \leftrightarrow \Box \Diamond \varphi \\
({U}_\Box) & & \square \square\varphi \leftrightarrow \square\varphi \leftrightarrow  \Diamond \square\varphi \\
\end{array}$\\
\end{proposition}

\begin{proof2}
It is easy to prove ${F}_{\Box\Diamond}$ using axioms $F_\Box$ and $D$. The details are left to reader.
For schemes $U_\Diamond$ and $U_\Box$,  axioms $4_\Box$, $4_\Diamond$, $5_\Box$ and $5_\Diamond$ give one direction of them. The opposite directions are obtained as follows:\\

\vspace{-0.1cm}
$
\begin{array} {l l l l}
\mbox{Proof 1:} & & \mbox{Proof 2:} & \\
\Diamond \varphi \to \Box \Diamond \varphi  & \mbox{ axiom } 5_\Diamond & \Diamond \Box\varphi \to \Box \varphi & \mbox{ axiom } 5_\Box \\
\Box(\varphi \to \Diamond\varphi) & \mbox{ by } MP \mbox{ and }  FS2   \hspace*{1cm} & \Box(\Box \varphi \to \varphi) & \mbox{ by } MP \mbox{ and }  FS2 \\
\Diamond \varphi \to \Diamond \Diamond \varphi & \mbox{ by } MP \mbox{ and }  P   & \Box \Box \varphi \to \Box \varphi & \mbox{ by } MP \mbox{ and }  K \\
\ &  \\
\mbox{Proof 3:} & & \mbox{Proof 4 :} & \\
 \Diamond \varphi \to \Diamond \varphi & \mbox{ prop. taut. } &  \Box \varphi \to \Box \varphi & \mbox{ prop. taut. } \\
 \Diamond( \Diamond \varphi \to \Diamond \varphi) & \mbox{ by } D &  \Diamond( \Box \varphi \to \Box \varphi) & \mbox{ by } D \\
\Box\Diamond \varphi \to \Diamond \Diamond \varphi & \mbox{ by } MP \mbox{ and } T5 & \Box\Box \varphi \to \Diamond \Box \varphi & \mbox{ by } MP \mbox{ and } T5 \\
\Box\Diamond \varphi \to  \Diamond \varphi & \mbox{ by } 4_\Diamond & \Box \varphi \to  \Diamond \Box \varphi & \mbox{ by } 4_\Box \\
\end{array}
$

\end{proof2}

From now on we will use $ThKD45(\mathbf{G})$ to denote the set of theorems of $KD45(\mathbf{G})$. We close this section with the following observation: deductions in $KD45(\mathbf{G})$ can be reduced to derivations in pure propositional G\"odel logic $G$.

\begin{lemma}
{\label{reduction} } For any theory $T$ and formula $\varphi $ in $\mathcal{L}_{\Box \Diamond
}$, it holds that $T\vdash _{KD45(\mathbf{G})} \varphi $ iff $T\cup ThKD45(\mathbf{G}) \vdash _{G} \varphi $.
\end{lemma}

\section{Possibilistic semantics and completeness}

In this section we will show that $KD45(\mathbf{G})$  is also complete with respect to the class of possibilistic G\"odel frames.

\begin{definition}
\label{simplgodelframe} A {\em possibilistic G\"odel frame} ($\Pi G$-frame) will be a
structure $\langle W,\pi \rangle $ where $W$\emph{\ is a non-empty
set of worlds, and }$\pi:W \rightarrow [0,1]$ is a normalized possibility distribution over $W,$ that is, such that $\sup_{w \in W} \pi(w) = 1.$

A {\em possibilistic G\"odel model} is a triple $ \langle W, \pi, e\rangle$ where $ \langle W, \pi\rangle $ is a $\Pi G$-frame frame and $e: W \times Var \to [0, 1]$ provides an evaluation of variables in each world.
For each $w \in W$, $e(w, \cdot)$ extends to arbitrary formulas in the usual way for the propositional connectives and for modal operators in the following way:

\medskip

$e(w,\Box \varphi ):=\inf_{w'\in W}\{\pi(w') \Rightarrow e(w',\varphi )\}$

$e(w,\Diamond \varphi ):=\sup_{w'\in W}\{\min(\pi(w'), e(w',\varphi ))\}$.
\end{definition}

Observe that the evaluation of formulas beginning with a modal operator is in fact independent from the current world.
As we already mentioned in the introduction, it is clear that a possibilistic frame $\langle W, \pi\rangle$ is equivalent to the GK-frame $\langle W, R_\pi \rangle$ where $R_\pi = W \times \pi$.

In the rest of the paper we provide a completeness proof of the logic $KD45(\mathbf{G})$ with respect of the class $\Pi\mathcal{G}$ of possibilistic G\"odel models, in fact we are going to prove weak completeness for deductions from finite theories.

In what follows, for any formula $\varphi$  we denote by $Sub(\varphi) \subseteq \mathcal{L}_{\square \Diamond }$
the set of subformulas of $\varphi$ and containing
the formula $\bot $.  Moreover, let $X:=\{\square \theta ,\Diamond \theta :\theta \in \mathcal{L}_{\square
\Diamond }\}$ be the set of formulas in $\mathcal{L}_{\square \Diamond }$
beginning with a modal operator; then $\mathcal{L}_{\square \Diamond }(Var)= 
\mathcal{L(}Var\cup X)$. That is, any formula in $\mathcal{L}_{\square
\Diamond }(Var)$ may be seen as a propositional G\"odel  formula built from the extended set
of  propositional variables $Var\cup X$. In addition, for a given formula $\varphi$, 
let $\sim_\varphi$ be equivalence relation in $[0,1]^{Var\cup X} \times \lbrack 0,1]^{Var\cup X}$
defined as follows:
$$ u \sim_\varphi w \mbox{ iff } \forall \psi \in Sub(\varphi): u(\Box \psi) = w(\Box \psi) \mbox{ and } u(\Diamond \psi) = w(\Diamond \psi) .$$

Now, assume that a formula $\varphi$ is not a theorem of $KD45(\mathbf{G})$. Hence by completeness of G\"odel calculus and Lemma \ref{reduction}, there exists a G\"odel valuation $v$  such that $ v(ThKD45(\mathbf{G}))=1$ and $v(\varphi)<1$.
Following the usual canonical model construction, once fixed the valuation $v$, we  define next a canonical $\Pi\cal G$-model $M^v_{\varphi}$ in which we will show $\varphi$ is not  valid.

The \emph{canonical model} \emph{\ }$
M^v_{\varphi}=(W^{v},\pi^{\varphi},e^{\varphi})$ is defined as follows:\medskip

\noindent $\bullet $ $W^{v}$ is the set $\{u \in \lbrack
0,1]^{Var\cup X} \mid u \sim_\varphi v \mbox{ and } u(ThKD45(\mathbf{G}))=1 \}$.

\noindent $\bullet $ $\pi^{\varphi}(u)=\min_{\psi \in Sub(\varphi)}\{\min(v(\Box \psi)\rightarrow
u(\psi ), u(\psi )\rightarrow v(\Diamond \psi ))\}.$

\noindent $\bullet $ $e^{\varphi}(u,p)=u(p)$ for any $p\in Var$.\medskip

In this context, we call the elements of $\Delta_\varphi = \{\square \theta ,\Diamond \theta :\theta \in Sub(\varphi) \}$, the {\em fixed points} of the Canonical Model.

Note that having $\nu(ThKD45(\mathbf{G}))=1$ does not guarantee that $\nu$ belongs to the canonical model because it may not take the appropriated values for the fixed points, i.e.\ it may be that  $v \not\sim_\varphi \nu$. However, the next lemma shows how, in certain conditions, to transform such an evaluation into another  belonging to the canonical model.

\begin{lemma}\label{normalization} Let $u \in W^v$ and let $\nu: {Var\cup X} \mapsto [0,1]$ be a G\"odel valuation. Define $\alpha = \max \{ u(\lambda) : \nu(\lambda) < 1 \mbox{ and } \lambda \in \Delta_\varphi\}$ and  $\beta = \min \{ u(\lambda) : \nu(\lambda)= 1 \mbox{ and } \lambda \in \Delta_\varphi\}$. If $\nu$ satisfies the following conditions: 
\begin{description}
\item [a.] $\nu(ThKD45(\mathbf{G}))=1$.
\item [b.] for any $\psi, \phi \in \{\lambda : u(\lambda) \leq \alpha \mbox{ and } \lambda \in \Delta_\varphi \}$, $\nu(\psi) < \nu(\phi)$ iff $u(\psi) < u(\phi)$.
\item [c.] $\nu(\lambda) = 1$ for every $\lambda \in \Delta_\varphi$ such that $u(\lambda)>\alpha$, 
\end{description}
then, there exists a G\"odel valuation $w \in W^v$ such that, for any formulas $\psi, \phi$:
\begin{enumerate}
\item $\nu(\psi) = 1 $  implies $ w(\psi) \geq \delta$.
\item $\nu(\psi) < 1 $ implies $ w(\psi) < \delta$.
\item $1 \neq  \nu(\psi) \leq \nu(\phi)$ implies $w(\psi) \leq w(\phi)$.
\item $\nu(\psi) < \nu(\phi)$ implies $ w(\psi) < w(\phi)$.
\item $\nu(\psi) = \nu(\phi) = 1  \mbox{ and } u(\psi) \leq u(\phi)$ imply $ w(\psi) \leq w(\phi)$.
\item $\nu(\psi) = \nu(\phi) = 1 \mbox{ and } u(\psi) < u(\phi)$ imply $ w(\psi) < w(\phi)$.
\end{enumerate}
\end{lemma}

\begin{proof2} See Appendix.
\end{proof2}

Completeness will follow from the next truth-lemma.

\begin{lemma} [Truth-lemma] \label{equation-joint} $e^{\varphi}(u,\psi )=u(\psi )$ for any $\psi \in
Sub(\varphi)$ and any $u\in W^{v}$.
\end{lemma}

\begin{proof2}
For simplicity, we write $W$ for $W^{v}.$ We prove the identity by
induction on the complexity of the formulas in $Sub(\varphi)$, considered now as elements
of $\mathcal{L}_{\square \Diamond }(Var)$. For $\bot $ and the propositional
variables in $Sub(\varphi)$ the equation holds by definition. The only non trivial
inductive steps are:\ $e^{\varphi}(u,\Box \psi)=u(\Box \psi)$ and $
e^{\varphi}(u,\Diamond \psi)=u(\Diamond \psi)$ for $\Box \psi,\Diamond
\psi \in Sub(\varphi).$ By the inductive hypothesis we may assume that $
e^{\varphi}(u^{\prime },\psi)=u^{\prime }(\psi)$ for every $u^{\prime }\in
W;$ thus we must prove
\begin{eqnarray}
\inf_{u^{\prime }\in W}\{\pi^\varphi(u^{\prime })\Rightarrow u^{\prime }(\varphi
)\}=u(\Box \varphi )  \label{box} \\
\sup_{v^{\prime }\in W}\{\min(\pi^\varphi(u^{\prime }), u^{\prime }(\varphi
)) \}=u(\Diamond \varphi )  \label{Diam}
\end{eqnarray}
By definition, $\pi^\varphi(u^{\prime })\leq (v(\Box \psi )\Rightarrow u^{\prime
}(\psi ))$ and $\pi^\varphi(u^{\prime })\leq (u^{\prime }(\psi )\Rightarrow
v(\Diamond \psi ))$ for any $\psi \in Sub(\varphi)$ and $u^{\prime }\in W;$
therefore, $u(\Box \psi) = v(\Box \psi) \leq (\pi^\varphi(u^{\prime })\Rightarrow u^{\prime
}(\psi))$ and $\min(\pi^\varphi(u^{\prime }), u^{\prime }(\psi)) \leq
v(\Diamond \psi) = u(\Diamond \psi).$ Taking  infimum over $u^{\prime }$ in the first
inequality and the supremum in the second we get
\begin{equation*}
u(\Box \psi )\leq \inf_{u^{\prime }\in W}\{\pi^\varphi(u^{\prime })\Rightarrow
u^{\prime }(\psi )\}, \ \sup_{u^{\prime }\in W}\{\min(\pi^\varphi(u^{\prime
}),  u^{\prime }(\psi ))\}\leq u(\Diamond \psi ).
\end{equation*}
Hence, if $u(\Box \psi )=1$ and $u(\Diamond \psi )=0$ we obtain (\ref
{box}) and (\ref{Diam}), respectively. Therefore, it only remains\ to prove
the next two claims for $\Box \psi ,\Diamond \psi \in Sub(\varphi)$.\medskip

\noindent \textbf{Claim 1}. \emph{If $u(\Box \psi )=\alpha <1$, for every $\varepsilon > 0$, there exists a valuation $w\in W$ such that $\pi^\varphi(w) > w(\psi)$ and $w(\psi ) < \alpha + \varepsilon$, and thus $(\pi^\varphi(w) \Rightarrow w(\psi ))< \alpha + \varepsilon$}.\medskip

\noindent \textbf{Claim 2. }\emph{If $u(\Diamond \psi )=\alpha >0$ then, for any $\varepsilon >0,$ there exists a valuation $w'\in W$ such
that $w'(\psi )=1$ and $\pi^\varphi(w')\geq \alpha -\varepsilon $, and thus $\min(w'(\psi ),  \pi^\varphi(w')) \geq \alpha -\varepsilon$}.

\medskip
The proof of these two claims are rather involved and they can be found  in  the appendix.
\end{proof2}

\begin{theorem} [Finite strong completeness]
\label{JointCompleteness} For any
finite theory $T$ and formula $\varphi $ in $\mathcal{L}_{\square \Diamond }$, $T\models _{\Pi\mathcal{G}}\varphi $ implies $T\vdash _{KD45(\mathbf{G})}\varphi .$
\end{theorem}

\begin{proof2} One direction is soundness, and it is easy to check that the axioms are valid in the class $\Pi\mathcal{G}$ of models.  As for the other direction, assume $T = \emptyset$ and $\not\vdash _{KD45(\mathbf{G})}\varphi .$ Then $ThKD45(\mathbf{G})\not\vdash _G\varphi $ by Lemma \ref{reduction}, and thus there is, by Proposition \ref%
{ordersoundness}, a G\"odel valuation $v:Var\cup X\rightarrow [0,1]$
such that $v(\varphi )< v(ThKD45(\mathbf{G}))=1.$ Then $v$ is a world of
the canonical model $M^\varphi_v$ and by Lemma \ref{equation-joint}, $e^{\varphi}(v,\varphi )=v(\varphi )<1.$ Thus $\not\models
_{\Pi \cal G}\varphi$. This proof can be easily generalized when $T$ is a non empty and finite.
\end{proof2}

\section{Conclusions}
In this paper we have studied the logic over G\"odel fuzzy logic arising from many-valued G\"odel Kripke models with possibilistic semantics, and have shown that it actually corresponds to a simplified semantics for the logic $KD45({\bf G})$, the extension of Caicedo and Rodriguez's bi-modal G\"odel logic with many-valued versions of the well-known modal axioms $D$, $4$ and $5$. The truth-value of a formula $\Diamond \varphi$ in a possibilistic Kripke model is indeed a proper generalization of the possibility measure of $\varphi$ when $\varphi$ is a classical proposition, however the semantics of $\Box \varphi$ is not. This is due to the fact that the negation in G\"odel logic is not involutive.

Therefore, a first open problem we leave for further research is to consider to extension of the logic $KD45({\bf G})$ with an involutive negation and investigate its possibililistic semantics. A second open problem is to investigate the logic arising from {\em non-normalized} possibilistic G\"odel frames. In the classical case, one can show that this corresponds to the modal logic $K45$, that is, without the axiom $D$, see e.g. \cite{Pietrus}. However, over G\"odel logic this seems to be not as straightforward as in the classical case.

\paragraph{Acknowledgments}
The authors are grateful to the anonymous reviewers for their helpful comments. They  acknowledge partial support by the H2020-MSCA-RISE-2015 project SYSMICS, the Spanish MINECO project RASO (TIN2015-71799-C2-1-P) and the Argentinean project PIP CONICET 11220150100412CO.

\section*{Appendix}

\noindent {\bf Proof of Lemma 2}

\begin{proof2}

First of all, notice that if $\nu$ satisfies the condition {\bf b}, then necessarily $\alpha < \beta$. 
Let  $B= \{ \nu(\lambda) : \lambda \in \Delta_\varphi , \nu(\lambda) < 1\} \cup \{0 \} = \{ b_0 = 0 < b_1 < \ldots b_N \}$. Obviously, $b_N < 1$. For each $0 \leq i \leq N$, pick  $\lambda_i \in \Delta_\varphi$ such that $\nu(\lambda_i) = b_i$.   
Define now a continuous strictly function $g:[0,1]\mapsto \lbrack 0,\delta) \cup \{1\}$ such
that \medskip

$g(1)=1$

$g(b_i)= v(\lambda_i)$ for every $0 \leq i \leq N$

$g[(b_N, 1)]= (\alpha, \delta)$
\medskip

\noindent Notice that $\alpha = g(b_N)$.  In addition, define another continuous strictly increasing function $h:[0,1] \mapsto [\delta, 1]$ such that
\medskip

$h(0)=\delta$

$h[(0,\beta)] =(\delta, \beta)$

$h(x) = x$, for $x \in [\beta, 1]$
\medskip

\noindent Then we define the valuation $w:  Var \cup X \to [0,1]$ as follows:

\medskip
$w(p) = \left \{ 
\begin{array}{ll} 
g(\nu(p)), & \mbox{if } \nu(p) < 1,\\
h(u(p)), & \mbox{if } \nu(p) = 1.
\end{array}
\right .
$
\medskip

\noindent First of all, let us show by induction that this extends to any propositional formula, that is, 

\medskip
$w(\varphi) = \left \{ 
\begin{array}{ll} 
g(\nu(\varphi)), & \mbox{if } \nu(\varphi) < 1,\\
h(u(\varphi)), &  \mbox{if } \nu(p) = 1.
\end{array}
\right .
$
\medskip

\noindent Note that, since  $g$ and $h$ are strictly increasing mappings, $g\circ\nu$ and $h \circ u$ are valuations as well. So, in the induction steps below we only need to check that everything is fine when both are used at the same time when  evaluating a compound formula. The base case holds by definition.

\begin{itemize}
\item Assume $\psi = \psi_1 \land \psi_2$. We only check the case when 
$v(\psi) < 1$ and $v(\psi_1) < 1$ and $v(\psi_2) = 1$. Then $w(\psi) = \min(w(\psi_1), w(\psi_2)) = \min(g(\nu(\psi_1)),  h(u(\psi_2)))  = g(\nu(\psi_1))$, since $g(\nu(\psi_1)) < \delta \leq h(u(\psi_2)$. But, $g(\nu(\psi_1)) = \min( g(\nu(\psi_1)),  1) =  \min( g(\nu(\psi_1)), g(\nu(\psi_2))) = g(\nu(\psi_1\land \psi_2)) = g(\nu(\psi))$. 

\item Assume $\psi = \psi_1 \to \psi_2$, and consider two subcases: 

(1)  $v(\psi_1) < 1$ and $v(\psi_2) = 1$. Then $v(\psi_1 \to \psi_2) = 1$ and  $w(\psi) = w(\psi_1) \Rightarrow w(\psi_2) = g(\nu(\psi_1)) \Rightarrow h(u(\psi_2)) = 1 =  h(u(\psi_1)) \Rightarrow h(u(\psi_2)) = h(u(\psi_1 \to \psi_2)) = h(u(\psi)). $

(2)   $v(\psi_1) = 1$ and $v(\psi_2) < 1$.  Then $v(\psi_1 \to \psi_2) = v(\psi_2) < 1$ and  $w(\psi) = w(\psi_1) \Rightarrow w(\psi_2) =  h(u(\psi_1))  \Rightarrow g(\nu(\psi_2)) = g(\nu(\psi_2))  = g(\nu(\psi_1)) \Rightarrow g(\nu(\psi_2)) = g(\nu(\psi_1 \to \psi_2)) = g(\nu(\psi))$. 
\end{itemize}

\noindent  Properties 1 -- 6 now  directly follow from the above. 

Finally, we prove that $w \in W^v$. By definition of $w$ it is clear that $w \sim_\varphi v$. It remains to check that $w$ validates all the axioms. The axioms of $\mathcal{G}$ are evaluated to $1$ by any G\"odel valuation. As for the specific axioms of $KD45(\mathbf{G})$, it is an immediate consequence of Property $3$ because it implies that if $\nu(\psi \to \phi)=1$ then $w(\psi \to \phi)=1$.
\end{proof2}

\medskip

\noindent \textbf{Claim 1 from Lemma \ref{equation-joint}}. \emph{If $u(\Box \psi )=\alpha <1$, for every $\varepsilon > 0$, there exists a valuation $w\in W$ such that $\pi^\varphi(w) > w(\psi)$ and $w(\psi ) < \alpha + \varepsilon$, and thus $\pi^\varphi(w) \Rightarrow w(\psi ) = w(\psi ) < \alpha + \varepsilon$}.\medskip

\begin{proof} By definition of G\"odel's implication $\Rightarrow $ in [0,1],
to grant the required conditions on $w$ it is enough to find $w \in W$  such that $\alpha \leq w(\psi)$ and, for any
$\theta \in Sub(\varphi)$, $u(\Box \theta )\leq w(\theta ) \leq u(\Diamond \theta ) \leq \alpha$.
This is achieved in two stages:
\begin{itemize}
\item first producing a valuation $ \nu \in W$ satisfying $\nu(\psi)<1$ and preserving the relative ordering conditions the values $w(\theta )$ must
satisfy, conditions which may be coded by a theory $\Gamma_{\psi,u}$;

\item and then moving the values $\nu(\theta )$, for $\theta \in Sub(\varphi)$,  to the correct
valuation $w$ by composing $\nu$ with an increasing bijection of [0,1].
\end{itemize}
Assume $u(\Box \psi)=\alpha <1$, and define (all formulas involved
ranging in\ $\mathcal{L}_{\square \Diamond }(Var)$)
\begin{equation*}
\begin{array}{ll}
\Gamma_{\psi,u}= & \{\lambda : \lambda \in \Delta_\varphi \mbox{ and } u(\lambda) > \alpha \} \\
& \cup \{\lambda \rightarrow \theta  : \lambda \in \Delta_\varphi \mbox{ and } u(\lambda) \leq u(\Box \theta) \} \\
& \cup \{(\theta \rightarrow \lambda)\rightarrow \lambda : \lambda \in \Delta_\varphi \mbox{ and }  u(\lambda)< u(\Box \theta)< 1 \} \\
& \cup \{\theta   \rightarrow \lambda : \lambda \in \Delta_\varphi \mbox{ and } u(\Diamond \theta) \leq u(\lambda) \} \\
& \cup \{(\lambda \rightarrow \theta)\rightarrow \theta : \lambda \in \Delta_\varphi \mbox{ and }  u(\Diamond \theta)< u(\lambda)< 1 \} \\
\end{array}%
\end{equation*}%
Then we have $u(\square \xi ) > \alpha $ for each $\xi \in \Gamma_{\psi,u}$. Indeed, first recall that, by $U_\Box$ and $U_\Diamond$ of Proposition \ref{simplif},  for any $\lambda \in  \Delta_{\varphi}$ we have $u(\lambda) = u(\Box \lambda) = u(\Diamond \lambda)$.  We analyse case by case. For the first set of formulas,  it is clear by construction. For the second set,  we have
$u(\square (\lambda \rightarrow \theta )) \geq u(\Diamond \lambda
\rightarrow \square \theta)= u(\Diamond \lambda) \Rightarrow u(\Box \theta) = u(\lambda) \Rightarrow u(\Box \theta) = 1$,  by $FS2$. For the third, by FS2 and P, we have $u(\square ((\theta \rightarrow \lambda)\rightarrow \lambda)) \geq u(\Diamond(\theta \to \lambda) \to \Box \lambda) \geq
u((\Box \theta  \to \Diamond \lambda) \to \Box \lambda) = 1$, since $ u(\Box\lambda) = u(\Diamond \lambda) = u(\square \theta \rightarrow \Diamond \lambda) < 1$. The fourth and fifth cases are very similar to the second and third ones respectively.

This implies
\begin{equation*}
\Gamma_{\psi,u} \not\vdash _{KD45(\mathbf{G})}\psi,
\end{equation*}
otherwise there would exist $\xi _{1},\ldots ,\xi _{k}\in \Gamma_{\psi,u}$
such that $\xi _{1},\ldots ,\xi _{k} \vdash _{KD45(\mathbf{G})} \psi.$ In such a case, we would have $\Box \xi _{1},\ldots ,\Box \xi _{k} \vdash _{KD45(\mathbf{G})}\Box \psi $ by $Nec$
and $K_{\square }$. Then $\Box \xi _{1},\ldots ,\Box
\xi _{k},ThKD45(\mathbf{G})\vdash _G\Box \psi$ by Lemma \ref{reduction} and thus by Proposition \ref{ordersoundness} (i), and recalling that $u(ThKD45(\mathbf{G}))=1,$
\begin{equation*}
\alpha < \inf u(\{\Box \xi _{1},\ldots ,\Box \xi _{k}\}\cup ThKD45(\mathbf{G}))\leq u(\square \psi )=\alpha ,
\end{equation*}
a contradiction. Therefore, by Proposition \ref{ordersoundness} (ii) there
exists a valuation $\nu:Var\cup X\mapsto \lbrack 0,1]$ such that $\nu(\Gamma_{\psi,u} \cup ThKD45(\mathbf{G}))=1$ and $%
\nu(\psi )<1$. This implies the following relations between $u$ and $\nu,$
that we list for further use. Given $\lambda \in \Delta_\varphi, \theta \in \mathcal{L}_{\square \Diamond }(Var) $, we have \label{proplambda}:\medskip
\begin{description}
\item[\#\textbf{1}.] If $u(\lambda )>\alpha $ then $\nu(\lambda)=1$
(since then $\lambda \in \Gamma_{\psi,u})$.

\item[\#\textbf{2}.] If $u(\lambda)\leq u(\square \theta)$  then $\nu(\lambda)\leq \nu(\theta )$ (since then $\lambda \rightarrow \theta \in \Gamma_{\psi,u})$. In particular, if $\lambda_1,\lambda_2 \in \Delta_\varphi$ and $u(\lambda_1)\leq u(\Box\lambda_2) = u(\lambda_2)$ then $\nu(\lambda_1)\leq \nu(\lambda_2)$. Furthermore,if $\Box \theta \in \Delta_\varphi$ then from $u(\Box \theta) = u(\Box\theta)$ by \#2, $\nu(\Box \theta) \leq \nu(\theta)$. That means, taking $\theta = \psi$,  $\nu(\Box \psi) \leq \nu(\psi) < 1$.

\item[\#\textbf{3}.] If $u(\lambda)< u(\square \theta)< 1$
then $\nu(\lambda)< \nu(\theta)$ or $\nu(\lambda)=1$ (since then $(\theta \rightarrow \lambda)\rightarrow \lambda)\in
\Gamma_{\psi,u}$). In particular, if $\lambda_1,\lambda_2 \in \Delta_\varphi$, $u(\lambda_1) < u(\lambda_2)$ and $u(\lambda_2) \leq u(\Box \psi)= \alpha $ then $\nu(\lambda_1) < \nu(\psi) < 1$ and thus $\nu(\lambda_1) < \nu(\lambda_2)$. This means that $\nu$ preserves in a strict sense the order values by $u$ of the formulas $\lambda \in \Delta_\varphi$ such that $u(\lambda) \leq \alpha$. 
\item[\#\textbf{4}.] If $ u(\Diamond \theta) \leq u(\lambda)$ then $\nu(\theta) \leq \nu(\lambda)$ (because $\theta \to \lambda \in \Gamma _{\psi ,u}$). In particular, if $\Diamond \theta \in \Delta_\varphi$ then $\nu(\theta) \leq \nu(\Diamond \theta)$.

\item[\#\textbf{5}.] If  $ u(\Diamond \theta)< u(\lambda)< 1$ then $\nu(\theta) < \nu(\lambda)$ or $\nu(\theta) =1$. In particular, if $\lambda_1, \lambda_2 \in \Delta_\varphi$ and $u(\lambda_1)< u(\lambda_2)\leq \alpha = u(\Box \psi)$ then $\nu(\lambda_1)< \nu(\lambda_2)$. Furthermore, if $u(\lambda_2)>0$ then $\nu(\lambda_2)>0$ (making $\lambda_1:=\Diamond\bot$ since $u(\bot )=u(\Diamond \bot)=0$).
\end{description}

According to the properties \#1, \#2 and \#3, it is clear that $\nu$ satisfies the conditions of Lemma \ref{normalization}.
Consequently, for all $\epsilon > 0$ (such that $ \alpha +\varepsilon<\beta$), taking  $\delta = \alpha + \varepsilon$  in Lemma \ref{normalization}, there exists a valuation $w \in W^{v}$ such that $w(\psi) < \alpha +\varepsilon = \delta$.
Then in order to finish our proof, it remains to show that:
\begin{equation}\label{A-bound}
  \pi^\varphi(w)= \inf_{\lambda \in sub(\varphi)} \min(v(\Box \lambda) \Rightarrow w(\lambda),  w(\lambda) \Rightarrow v(\Diamond \lambda)) >w(\varphi)
\end{equation}
To do so, we will prove that, for any $\lambda \in sub(\varphi)$, both implications in (3) are  greater than $\delta$. \footnote{Remember that $u \sim_\varphi v \sim_\varphi w$.}
First we prove it for the first implication by cases:
\begin{description}
\item[-] If $\ v(\Box \lambda ) \leq \alpha < 1$ then $\min(v(\Box \lambda) \Rightarrow w(\lambda), w(\lambda) \Rightarrow v(\Diamond \lambda)) = 1$. Indeed, first of all, by   \#2,  
from $u (\Box \lambda) = v(\Box \lambda ) \leq \alpha = u(\Box \psi)$ it follows $\nu(\Box \lambda) \leq \nu(\psi) < 1$. Now, since $u(\Box\lambda) \leq u(\Box\lambda)$, by \#2, we have $1 \neq \nu(\Box \lambda) \leq \nu(\lambda)$, and by 3 of Lemma \ref{normalization} we have $ v(\Box\lambda) = w(\Box \lambda) \leq w(\lambda)$. Then $v(\Box \lambda) \Rightarrow w(\lambda) =1$.

\item[-] If $v(\Box \lambda ) > \alpha$  then by \#1 and \#2, $1=\nu(\Box \lambda)\leq\nu(\lambda)$. Therefore, by 1 of Lemma \ref{normalization}, $w(\lambda)>\delta$ 
which implies $v(\Box\lambda) \Rightarrow w(\lambda) > \delta$. 
\end{description}

\noindent For the second implication we also consider  two cases:
\begin{description}
\item[-] If $v(\Diamond \lambda) = u(\Diamond \lambda) > \delta$ then it is obvious that $w(\lambda) \Rightarrow v(\Diamond \lambda) > \delta$.

\item[-]  If $u(\Diamond \lambda) < \delta$, by definition of $\delta$ and taking into account that $\Diamond \lambda \in \Delta_\varphi$, then $u(\Diamond \lambda) < \alpha$. Now from $u(\Diamond \lambda) = u(\Diamond \lambda)$ we obtain by \#4, that  $\nu(\lambda) \leq \nu(\Diamond \lambda) < 1$. Then by Lemma \ref{normalization} we have $w(\lambda) \leq w(\Diamond \lambda) = v(\Diamond \lambda)$ and thus $w(\lambda) \Rightarrow v(\Diamond \lambda) = 1$.  \hfill $\blacksquare$
\end{description}
\end{proof}
\vspace{0.2cm}

\noindent \textbf{Claim 2 from Lemma  \ref{equation-joint}.  }\emph{If $u(\Diamond \psi )=\alpha >0$ then, for any $\varepsilon >0,$ there exists a valuation $w'\in W$ such
that $w'(\psi )=1$ and $\pi^\varphi(w')\geq \alpha -\varepsilon $, and thus $\min(w'(\psi ), \pi^\varphi(w')) \geq \alpha -\varepsilon$}.

\begin{proof2}
Assume $u(\Diamond \psi )=\alpha >0$ and define $\Gamma_{\psi,u}$ in the same way that it was defined in the proof of Claim 1.
Then we consider two cases:
\begin{description}
\item[- If] $u(\Diamond \psi )=1$,  let $U_{\psi,u}^{{}} =  \{ \lambda  : \  \lambda \in \Delta_\varphi \mbox{ and }  u(\lambda) < 1 \}$. We claim that
\end{description}
\begin{equation*}
\psi , \Gamma _{\psi,u} \not\vdash _{KD45(\mathbf{G})}  \bigvee U_{\psi ,u} \ ,
\end{equation*}
otherwise we would have $\theta_1, \ldots, \theta_n \in \Gamma _{\psi,u}$ such that $\vdash _{KD45(\mathbf{G})} \psi \rightarrow ((\theta_1\wedge \ldots \wedge \theta_n) \to \bigvee U_{\psi ,u})$,  and then we would also have $\vdash _{KD45(\mathbf{G})} \Diamond \psi \rightarrow \Diamond ((\theta_1\wedge \ldots \wedge \theta_n)  \to \bigvee U_{\psi ,u})$, that would imply in turn that
$\vdash _{KD45(\mathbf{G})} \Diamond \psi \rightarrow ((\Box\theta_1\wedge \ldots \wedge \Box\theta_n)  \to \Diamond\bigvee U_{\psi,u})$.
In that case, taking the evaluation $u$ it would yield: $ 1= u(\Diamond \psi)  \leq u(\Box \theta_1\wedge \ldots \wedge \Box\theta_n ) \Rightarrow u( \Diamond\bigvee U_{\psi ,u})$, a contradiction, since $u(\Box\theta_1\wedge \ldots \wedge \Box\theta_n) = 1$ and $u( \Diamond\bigvee U_{\psi ,u})) < 1$.\footnote{Note that, in this case, the first subset of $\Gamma _{\psi,u}$ is empty.}

Therefore, there is a G\"odel valuation $\nu'$ (not necessarily in $W$) such that $\nu'(\psi)=\nu'(\Gamma  _{\psi,u})=\nu'(ThKD45(\mathbf{G}))=1$ and $\nu'(\bigvee U_{\psi ,u} )<1$. By \#2 and \#3, it follows that for any $\lambda_{1},\lambda_{2} \in \Delta_\varphi$ such that $u(\lambda_{1}),  u(\lambda_{2}) \leq \alpha$,  we have  
$u(\lambda_{1})< u(\lambda_{2}) \leq \alpha \ \mbox{ iff } \ \nu'(\lambda_{1}) < \nu'(\lambda_{2}). $
Thus, $\nu'$ satisfies the conditions of Lemma \ref{normalization} because it is strictly increasing in $\Delta_\varphi$ (i.e. it satisfies condition {\bf b} of Lemma  \ref{normalization}), and $\nu'(ThKD45(\mathbf{G}))=1$. Therefore, there exists a valuation $w' \in W$ such that $w'(\psi)=1$.

 It remains to show that $\pi^\varphi(w') = 1$.
Indeed, by construction, it holds that $u(\Box \theta) \leq w'(\theta )\leq u(\Diamond \theta)$, and hence $\min(u(\Box \theta) \Rightarrow w'(\theta ) , w'(\theta )\Rightarrow u(\Diamond \theta )) =1$. 

\begin{description}
\item[- If] $1> u(\Diamond \psi )=\alpha >0$, then we let $U_{\psi,u}^{{}} = \ (\Diamond \psi \to \psi) \to \psi .$ We claim that
\end{description}
\begin{equation*}
\Box \top , \Gamma _{\psi,u} \not\vdash _{KD45(\mathbf{G})} U_{\psi ,u} \ ,
\end{equation*}
otherwise there would exist $\theta_1, \ldots, \theta_n \in \Gamma _{\psi,u}$ such that $\vdash _{KD45(\mathbf{G})} \Box \top \rightarrow ((\theta_1\wedge \ldots \wedge \theta_n) \to U_{\psi ,u})$, and then we would have $\vdash _{KD45(\mathbf{G})} \Diamond \Box \top \rightarrow \Diamond ((\theta_1\wedge \ldots \wedge \theta_n)  \to U_{\psi ,u})$, which would imply
$\vdash _{KD45(\mathbf{G})} \Box \top \rightarrow ((\Box\theta_1\wedge \ldots \wedge \Box\theta_n)  \to \Diamond U_{\psi,u})$.
In that case, evaluating with $u$ it would yield $ 1= u(\Box \top)  \leq u(\Box \theta_1\wedge \ldots \wedge \Box\theta_n ) \Rightarrow u( \Diamond U_{\psi ,u}))$, contradiction, since $u(\Box\theta_1\wedge \ldots \wedge \Box\theta_n) > \alpha$  and $u( \Diamond U_{\psi ,u})) \leq \alpha$ (because $u(\Diamond((\Diamond \psi \to \psi) \to \psi))\leq u(\Box(\Diamond \psi \to \psi) \to \Diamond \psi)\leq u(\Diamond \psi)\leq \alpha$).

Therefore, there is an evaluation $\nu'$ such that $\nu'(ThKD45(\mathbf{G}) )=\nu'(\Gamma  _{\psi,u})=1$ and $\nu'(\bigvee U_{\psi ,u} )<1$. Hence, we can conclude that the three pre-conditions $\bf a$, $\bf b$ and $\bf c$ required in Lemma \ref{normalization} are satisfied. 
In addition, the following condition is also satisfied:
\begin{description}
\item[d.]  $\nu'(\Diamond \psi) = \nu'(\psi)$.
\end{description}
At this point, we can now do a  proof dual to the one for Claim 1. Again, by Lemma \ref{normalization} for $\delta = \frac{\beta - \alpha}{2}$, we obtain from $\nu'$ an evaluation $w' \in W^{v}$ such that $w'(\psi)=\alpha$.
It only remains then to show that $\pi^\varphi(w)> \alpha$. But in this case, the proof is the same than the one given for equation (\ref{A-bound}) using $w'$ instead of
$w$. This finishes the proof.
\end{proof2}
\end{document}